# Highly efficient spin transport in epitaxial graphene on SiC

Bruno Dlubak[1], Marie-Blandine Martin[1], Cyrile Deranlot[1], Bernard Servet[2], Stéphane Xavier[2], Richard Mattana[1], Mike Sprinkle[3], Claire Berger[3,4], Walt A. De Heer[3], Frédéric Petroff[1], Abdelmadjid Anane[1], Pierre Seneor[1] and Albert Fert[1]

1. Unité Mixte de Physique CNRS/Thales, 91767 Palaiseau, France associée à l'Université de Paris-Sud 11, 91405 Orsay, France,

2. Thales Research and Technology, 91767 Palaiseau, France,

3. School of Physics, Georgia Institute of Technology, Atlanta, Georgia 30332, USA,

4. CNRS-Institut Néel, 38042 Grenoble, France.

**Spin information processing is a possible new paradigm for post-CMOS (complementary metal-oxide semiconductor) electronics and efficient spin propagation over long distances is fundamental to this vision. However, despite several decades of intense research, a suitable platform is still wanting. We report here on highly efficient spin transport in two-terminal polarizer/analyser devices based on high-mobility epitaxial graphene grown on silicon carbide. Taking advantage of high-impedance injecting/detecting tunnel junctions, we show spin transport efficiencies up to 75%, spin signals in the mega-ohm range and spin diffusion lengths exceeding 100 µm. This enables spintronics in complex structures: devices and network architectures relying on spin information processing, well beyond present spintronics applications, can now be foreseen.**

In 1990, the idea of a spin transistor based on spin transport in a semiconducting channel was introduced[1]. Since then, injecting spins (binary states of electrons, spin up or spin down) into lateral nanodevices, transforming spin information into electrical output signals and manipulating these signals with a gate have been persistent problems[2-8]. With conventional semiconductors[6-8] and metals[5], in spite of sustained efforts, only modest output signals were obtained[9-11]. The spin signals $\Delta R$ (that is, the absolute magnetoresistances, which correspond to the difference in the resistance measured when magnetizations of the ferromagnetic electrodes are parallel and antiparallel) do not generally exceed a few ohms and the spin diffusion length characterizing the lateral propagation of spin information hardly exceeds 1 µm. On the contrary, targeting spin information processing devices and architectures, both a large $\Delta R$ spin signal and a large MR (the relative magnetoresistance, that is the spin signal normalized by the device resistance) are required, to discriminate them from background and spurious effects and to avoid unpractical four-terminal device geometries. More recently, carbon-based materials have emerged as promising candidates for spin information transport. In these materials, long spin lifetimes are anticipated because of the small spin-orbit coupling and hyperfine interactions[12]. Whereas large signals have been observed in carbon nanotubes[13], only relatively small spin signals have been reported in graphene-based devices[14-19]. Consequently, graphene's potential is still to be unleashed.

Here we show measurements of highly efficient spin information transport and very large spin signals in spintronic devices patterned on epitaxial graphene that was grown on a 4 mm × 5 mm SiC chip (Fig. 1). We focus on measurements performed at 2 K of several two-terminal devices consisting of $w$ = 10-µm-wide graphene channels connected to Co electrodes through $Al_2O_3$ tunnel barriers. The device parameters are extracted through the analysis of the variation of the MR with the electrode/epitaxial graphene contact resistance $R_b$ and channel length $L$. The epitaxial graphene samples were grown by confinement-controlled sublimation of Si atoms on the C face of a SiC substrate[20]. This material was chosen because it allows large-scale integration without compromising transport properties. Epitaxial graphene has demonstrated mobilities up to 27,000 cm$^2$V$^{-1}$s$^{-1}$ (ref. 21) (17,000 cm$^2$V$^{-1}$s$^{-1}$ in this work) at a charge density of a few $10^{12}$ cm$^{-2}$ and high Fermi velocities of $1.0 \times 10^6$ ms$^{-1}$ (ref. 22). The epitaxial graphene layer is grown on the carbon-terminated silicon carbide face and is composed of a few rotationally stacked monolayer graphene sheets (however, it specifically is not a thin graphite layer)[22,23].

The output spin signal of devices composed of a non-magnetic lateral channel (here graphene) between ferromagnetic electrodes can be expressed in terms of the spin-up ($\sigma = \uparrow$) and spin-down ($\sigma = \downarrow$) contact resistances, $R_{b\sigma} = 2(1 \pm \gamma)R_b$, and the spin resistance of the channel, $R_{ch}^s = \rho l_{sf}/w$ (corresponding to the resistance experienced by the spin during its lifetime while dwelling in the channel), where $\gamma$ is the spin polarization of the tunnel barrier (that is, the asymmetry of spin resistances), $\rho$ is the resistivity per square of the channel and $l_{sf}$ its spin diffusion length[24,25]. The spin signal is governed by the ratio of the spin relaxation rate in the channel (which is inversely proportional to $R_{ch}^s$) to the spin relaxation rate due to the escape of spin currents into the ferromagnetic electrodes (which is inversely proportional to $R_b$, here a tunnel resistance). The standard expressions reported in the literature[24] apply to the limit where spin relaxation occurs mainly in the channel, so that the spin signal is determined (and limited) by $R_{ch}^s$. However, when the channel is a carbon-based material having a very slow spin relaxation, much larger spin signals can be obtained when the spin relaxation is mainly controlled by spin escape into the electrodes and inversely proportional to $R_b$ (refs 11,25). Specifically, in the case that $R_{ch}^s < R_b < R_{ch}^s\ d$ (where $d = L\ l_{sf}$) the spin signal is set by the tunnel resistance[25] and can hence become much larger than in the previous limit (one tends to the ideal situation of a double magnetic tunnel junction with zero spin relaxation between the two tunnel barriers). We show here that large output spin signals comparable to the tunnel resistances can be obtained with epitaxial-graphene-based lateral devices.

We next present resistance versus magnetic field measurements on a set of two-terminal devices and first focus on two devices of the set (Fig. 2). In device A (Fig. 2a) the distance between the electrodes is $L = 2$ µm and the total resistance is $R_P = 136$ MΩ in the configuration with parallel magnetizations of the electrodes. For device E (Fig. 2b) $L = 0.8$ µm and $R_P = 5.85$ MΩ. Device A yields $\Delta R = 1.5$MΩ for the difference between the resistances of the parallel and approximately antiparallel magnetic configurations whereas $\Delta R = 0.55$ MΩ for device E. The spin signals are in the mega-ohm range, which is up to four orders of magnitude greater than reported for other graphene devices[14-19]. This very large spin signal cannot be explained by spurious effects originating from the channel or contacts (see later) or by the conventional model where $\Delta R$ is typically found to be comparable to the channel spin resistance $R_{ch}^S$ (ref. 24). Hence, with $\rho \approx 1$ KΩ (inferred from Hall bar measurements), $w = 10$ µm, $R_{ch}^S$ expressed in kilo-ohms and $l_{sf}$ in micrometres, one obtains $R_{ch}^S \approx l_{sf}\ 10$, so that to account for values of $\Delta R$ in the mega-ohm range by the standard equations[24] where $\Delta R \approx R_{ch}$ would require too unrealistic values of $l_{sf} = 10$ mm. Our large values of $\Delta R$ necessarily imply that the scale of $\Delta R$ is set by the large resistance $R_b$ of the tunnel barrier contacts[25]. As the dominant contribution to the resistance of our devices comes from the tunnel junctions, the contact resistance $R_b$ can be expressed as $R_b = R_P/2(1 - \gamma^2)$, where $R_P$ is the total device resistance in the parallel state if, for simplicity, we assume the same values for the spin polarization $\gamma$ and $R_b$ of the two electrodes. An equation equivalent to equation (1) for a device with two different tunnel resistances for the two electrodes can be derived easily from the formalism of ref. 25 but we have preferred using equation (1), which is simpler and allows us a more transparent discussion of the essential physics. We have checked that an analysis with different tunnel resistances in proportion to their areas leads to practically the same conclusions. Then the standard drift/diffusion equations lead to the following expression[25] of $\Delta R$ in the limit of negligible ferromagnet spin resistance $R_F \ll R_b$:

$$\Delta R = \frac{4\gamma^2 R_b}{2\cosh\frac{L}{l_{sf}} + \left(\frac{R_b}{R_{ch}^S} + \frac{R_{ch}^S}{R_b}\right)\sinh\frac{L}{l_{sf}}} \qquad (1)$$

If $R_b \gg R_{ch}^S$ and $l_{sf} \gg L$ this can be written as:

$$\Delta R = \frac{2\gamma^2 R_b}{1 + \frac{L}{2l_{sf}}\frac{R_b}{R_{ch}^S}} \qquad (2)$$

For small $R_b$, $\Delta R$ is proportional to $R_b$; for $R_b \gg 2R_{ch}^S l_{sf}\ L$ it saturates at $\Delta R = 4\gamma^2 R_{ch}^S l_{sf}\ L$. In Fig. 3, we show the trend of $\Delta R$ as a function of the tunnel barrier resistance $R_b$ for typical parameters of our samples ($L = 2$ µm, $w = 10$ µm), a typical value of $\gamma = 0.32$ for Co/alumina junction and three different $l_{sf}$. Consequently, to observe $\Delta R$ spin signals in the mega-ohm range requires not only very large $l_{sf}$ (in the 100 µm range and above) but also highly resistive injector/detector barriers. Moreover, if both conditions are not simultaneously met, $\Delta R$ is limited to a few kilo-ohms. The very long spin diffusion length $l_{sf}$ of the order of 100 µm in the graphene of our devices is confirmed in the analysis below.

We next focus on the magnitude of the MR. Table 1 lists the geometrical parameters, the barrier resistances $R_b$ and the spin signals for the set of five devices (A-E). The largest MR (9.4%) is obtained for the smallest resistance (device E, also see Fig. 4). This value should be compared with the maximum achievable MR with a perfect channel: that is 12.5%. A perfect channel gives 1/2 of the MR of a Co/Al$_2$O$_3$/Co magnetic tunnel junction[26] (see equation (4)), which amounts to 25% in our devices[27]. Consequently, the MR = 9.4% of device E translates into a 75% overall device efficiency. This implies an exceptionally small degradation of spin information during the transport through the graphene channel. As $R_b$ increases, the MR quickly drops into the few per cent ranges as expected from the following equation derived from equation (1) and $R_P = 2R_b(1-\gamma^2)$:

$$MR = \frac{\Delta R}{R_P} = \frac{\gamma^2}{1-\gamma^2} \frac{2}{2\cosh\frac{L}{l_{sf}} + \left(\frac{R_b}{R_{ch}^S} + \frac{R_{ch}^S}{R_b}\right)\sinh\frac{L}{l_{sf}}} \quad (3)$$

We used equation (3) to derive the spin diffusion length $l_{sf}$ for each sample from its MR. We note here that the only free parameters are $l_{sf}$ and $\gamma$. In the following calculation we conservatively used $\gamma = 0.32$, which is the maximum value measured in our tunnel junctions[27]. From Table 1, devices B-D exhibit similar $l_{sf}$ in the 100-200 µm range and devices A and E have a larger value of $l_{sf}$ in the 250 µm range. Typically $l_{sf}$ is $\geq$ 150 µm and we ascribe the degraded properties of devices B-D to steps or domain edges crossing the channel. These values of $l_{sf}$ are two orders of magnitude higher than the previously reported $l_{sf}$ for graphene layers or multilayers and confirm the potential of graphene and graphene on SiC in particular for spintronics[28]. Note that smaller values of $\gamma$ would lead to longer $l_{sf}$, whereas $\gamma = 0.4$, the largest observed $\gamma$ for tunneling from Co through alumina, would lead to slightly shorter $l_{sf}$, for example 108 µm instead of 138 µm for sample C. We now focus on the variation of the MR with $R_b$ and $L$. As $l_{sf} \gg L$ we can rewrite equation (3) as a function of $R_b \cdot L$:

$$\Delta R = \frac{\gamma^2}{1-\gamma^2} \frac{1}{1+\frac{L}{2l_{sf}}\frac{R_b}{R_{ch}^S}} \quad (4)$$

Figure 4 plots equation (4) for $l_{sf} = 4$ µm, $l_{sf} = 50$ µm, $l_{sf} = 150$ µm and $l_{sf} = 250$ µm. The measured MR versus $R_b \cdot L$ data for devices A-E are also shown. Figure 4 shows that equation (4) captures the experimental trends for the dependence of $R_b$ and $L$ and also allows to clearly discriminate between the various values of $l_{sf}$. Consequently we need not rely on the Hanle effect, which neglects electrode effects and is known to fail for resistive contacts[19,29]. This is particularly true for highly efficient spin transport reported here. In our case, the spin relaxation rate measured by the Hanle effect represents the escape rate through the contacts rather than an intrinsic channel property. Our simple resistance model readily takes this into account as evidenced by the remarkable quantitative agreement regarding the decrease of the MR with $R_b \cdot L$ (equation (2)). As $R_b$ increases, both $\Delta R$ and the total resistance $R_P$ first increase proportionally to $R_b$, then, for $R_b = 2R_{ch}^s l_{sf} \ L$, $\Delta R$ saturates whereas $R_P$ continues to increase, so that MR decreases as seen experimentally. Further increasing $R_b \cdot L$ increases the dwell time $t_n$ of the carriers in the channel, eventually causing loss of the carried spin information thereby reducing the efficiency. The denominator of equation (4) can also be written as $1 + t_{sf} \ t_n$, where $t_{sf}$ is the spin lifetime[11], which leads to the expression used for the interpretation of results on carbon nanotubes[13].

One could argue that the observed magnetoresistive signal does not originate from spin transport but from spurious effects in the graphene itself or at the contacts. On one hand, the fact that the effect could originate from spurious effects in the graphene channel or the electrodes can be ruled out as its resistance (kilo-ohm range) is several orders of magnitude below the observed effect one (mega-ohm range). On the other hand, the observed effect cannot be ascribed to the anisotropic MR of the contacts alone (such as tunnel anisotropic MR for example). Indeed, this could not explain the observed decrease of the MR (and even less the hyperbolic $1/x$ behaviour) with the increase of the contact resistance (Fig. 4). On the contrary, the hyperbolic behaviour is very well described in our conventional model of spin transport.

The very high $\Delta R$ spin signals of our samples rely on the large $l_{sf}$ in epitaxial graphene. The nearly two orders of magnitude greater $l_{sf}$ compared with previously reported values can be ascribed to several factors. Previously reported $l_{sf}$ may have been underestimated in the case of lower-quality tunnel barrier contacts or increased interface roughness as recently evidenced[19,30]. More importantly, compared with $SiO_2$-supported graphene layers, the multilayer nature of epitaxial graphene allows two main advantages: the first one is the screening of the substrate-induced scattering leading to reduced spin relaxation through the Elliot-Yafet mechanism[15]. The reduction of the scattering in our graphene is also confirmed by the high mobility in the range of 17,000 $cm^2V^{-1}s^{-1}$ measured in our Hall bar samples. The second point is that epitaxial graphene is much flatter than monolayer graphene exfoliated on $SiO_2$ (refs 31, 32). As already expressed for exfoliated bilayers and multilayers[33], the smoothing out of the graphene sheets is thought to reduce the ripple-induced large spin-orbit coupling and spin relaxation through the Dyakonov-Perel mechanism[12]. From the observation of $l_{sf} > 100$ µm and a mean free path of ~0.1 µm, we deduce a spin lifetime > 100 ns. We believe that this high mobility and the absence of ripples are key elements towards unlocking the spin transport properties of our samples or other graphene-based devices, although we acknowledge that a unified picture of spin relaxation in these materials is still missing.

These experiments demonstrate the potential of graphene for spin information processing. Large spin signals were propagated and detected in two-terminal devices. Spin transport efficiencies were measured to be as large as 75% of that expected for a perfect transmission channel. The large spin signals and large device impedances were shown to scale to, and be driven by, the adaptable tunnel barriers. In combination with wafer-scale epitaxial graphene, this paves a way for spin information processing involving large scale spin-based logic devices and networks. It is noteworthy that the best available tunnel barriers give rise to tunnel MR (TMR) signals >1,000% (ref. 34). This means that lateral devices with MR = 500% (TMR/2) could be obtained for an ideal channel. Hence, with a spin transport channel of high efficiency up to 75%, spin logic arrays in a wide adaptable range of device resistances could be realized with up to MR = 400% signals (75% of TMR/2).

# Methods

Device fabrication. The device fabrication is based on scalable standard lithography processes, either ultraviolet (>1 µm features) or electron-beam (<1 µm features, Nanobeam nB3 lithography system). First, graphene channel stripes with $w = 10$ µm width are defined by $O_2$ plasma etching. These stripes are then contacted with gold-capped $Al_2O_3$(1 nm)/Co(15 nm) spin injector/detector electrodes grown by sputtering. The two magnetic electrodes have different widths, which results in different magnetization reversal fields and allows us to measure the difference in the response between the parallel (P) and antiparallel (AP) orientations of the electrode's magnetizations. The $Al_2O_3$ tunnel barrier is deposited in two steps: a 0.6 nm Al layer is sputtered in 2 $Wcm^{-2}$ Ar d.c. plasma and then further oxidized in a 50 torr $O_2$ atmosphere. This standard procedure leads to high-quality 1-nm-thick tunnel barriers. The flatness (r.m.s. < 0.2 nm, no pinholes) of this barrier on graphene and the homogeneity of its electrical properties (no hotspots) were checked by atomic force microscopy and conductive-tip atomic force microscopy measurements (not shown). The electrical measurements confirm the high quality of the tunnel barriers with resistances in the mega-ohm range and resistance×area products greater than 1 $M\Omega.\mu m^2$. Resistance and Hall effect measurements have been used to derive a carrier density of $5\times10^{12}$ $cm^{-2}$ and a mobility around 17,000 $cm^2V^{-1}s^{-1}$.

# Table 1 – Devices parameters

| Device | A | B | C | D | E |
|---|---|---|---|---|---|
| Inter-electrode distance $L$ | 2 µm | 2 µm | 2 µm | 2 µm | 0.8 µm |
| Device resistance | 136 MΩ | 70 MΩ | 29 MΩ | 3.8 MΩ | 5.8 MΩ |
| Barrier resistance $R_b$ | 75.8 MΩ | 39 MΩ | 16.2 MΩ | 2.1 MΩ | 3.2 MΩ |
| Measured spin signal $\Delta R$ | 1.5 MΩ | 0.4 MΩ | 0.35 MΩ | 0.12 MΩ | 0.55 MΩ |
| Measured MR | 1.1% | 0.7% | 1.2% | 3.4% | 9.4% |
| $l_{sf}$ | 285 µm | 160 µm | 138 µm | 95 µm | 246 µm |

Measured signals and characteristics for devices A to E together with the $l_{sf}$ derived from equation (3) with $w = 10$ µm, $\rho = 1$ KΩ and $\gamma = 0.32$. We note here that, as we typically expect in our devices to range from 0.3 to 0.32, the value $\gamma = 0.32$ we chose gives us a lower bound for $l_{sf}$.

# Figures captions

**Figure 1 - Devices patterned on epitaxial graphene.**

**a**, Scanning electron micrograph of a two-terminal lateral spin valve with a distance $L = 2$ μm between the Al$_2$O$_3$/Co electrodes (coloured in red) deposited on the $w = 10$-μm-wide epitaxial graphene (EG) channel grown on the C face of the SiC substrate (coloured in blue). **b**, Optical image of the set of two-terminal spintronics devices (left) and of a Hall bar device (right), both built on the same epitaxial graphene sheet. **c**, Sketch representing the device geometry.

**Figure 2 - Magnetotransport measurements.**

**a,b**, Resistance versus magnetic field data recorded at $T = 1.4$ K and 20 mV on two devices presenting the largest $\Delta R$ (**a**) and the largest MR (**b**) of our set. Device characteristics are: (**a**) $L = 2$ μm, $R_P = 136$ MΩ (resistance in the parallel magnetization state), $\Delta R = 1.5$ MΩ, MR ~ 1% (device A in Table 1), (**b**) $L = 0.8$ μm, $R_P = 5.85$ MΩ, $\Delta R = 0.55$ MΩ, MR = 9.4% (device E in Table 1).

**Figure 3 - Achievable spin signal amplitude $\Delta R$.**

Calculation of the signal amplitude $\Delta R$ as a function the tunnel barrier resistance $R_b$ following equation (1) for $l_{sf} = 50$ μm, $l_{sf} = 150$ μm and $l_{sf} = 250$ μm. The graph shows the combined requirement of long $l_{sf}$ and mega-ohm range $R_b$ to achieve large spin signals approaching the mega-ohm range. The other parameters, inferred from our samples, are $L = 2$ μm, $w = 10$ μm and $\rho = 1$ KΩ; $\gamma$ is set to 0.32.

**Figure 4 - Consistency of measured spin signals with standard transport models.**

Evaluated MR calculated from equation (4) and expressed as a function of the $R_b \cdot L$ product for a $l_{sf}$ of 4 μm (literature), 50 μm, 150 μm and 250 μm and $w = 10$ μm, $\rho = 1$ KΩ and $\gamma = 0.32$. Symbols: measured MR of devices shown in Table 1.

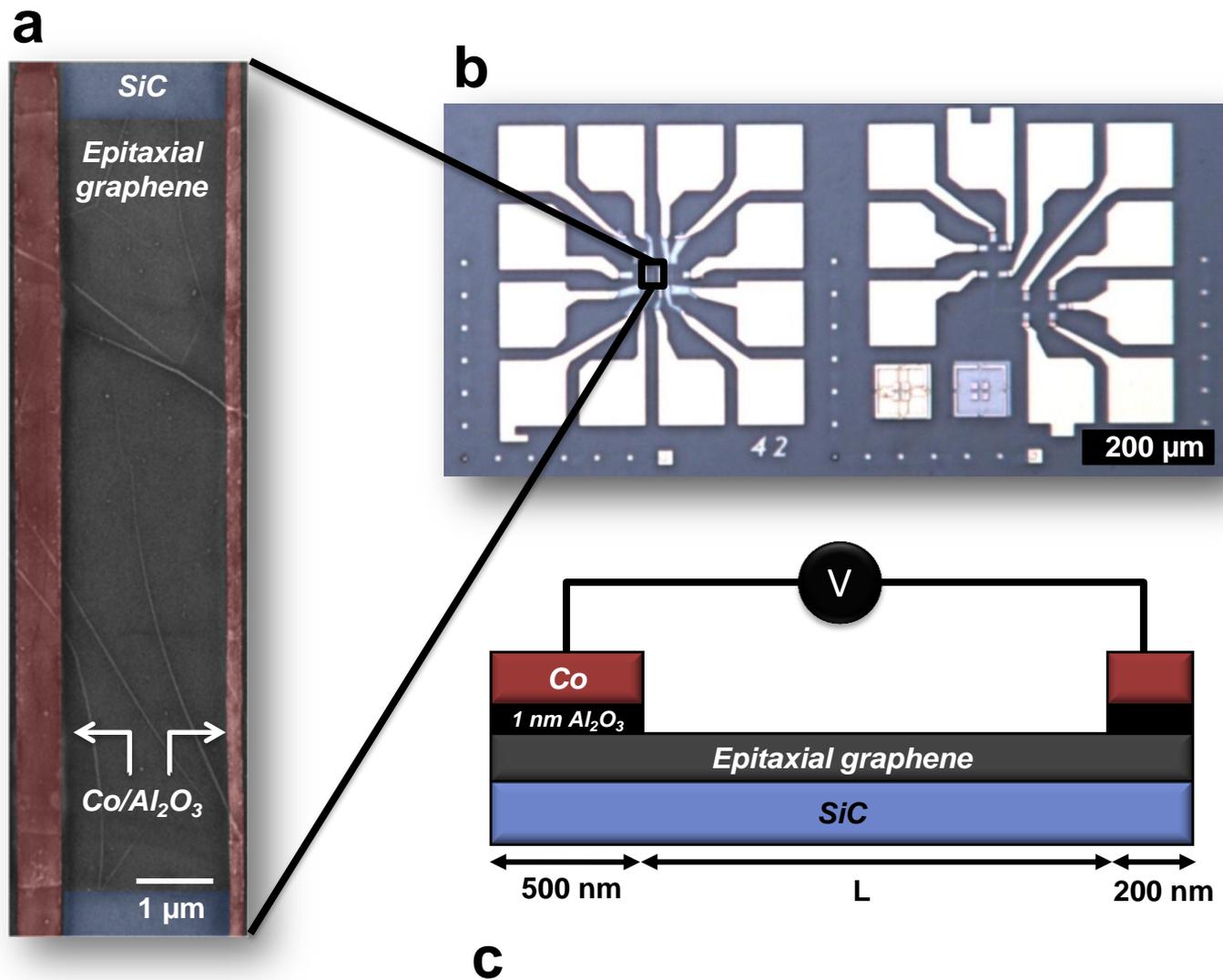

**Figure 1**

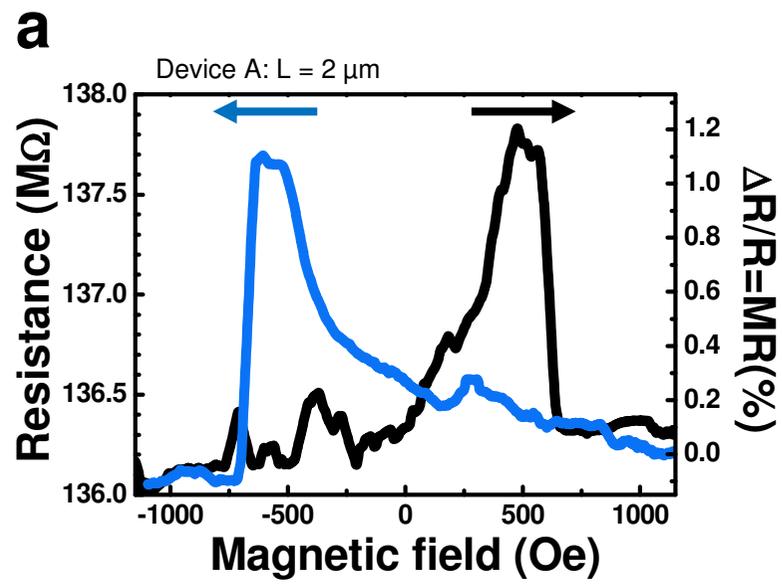 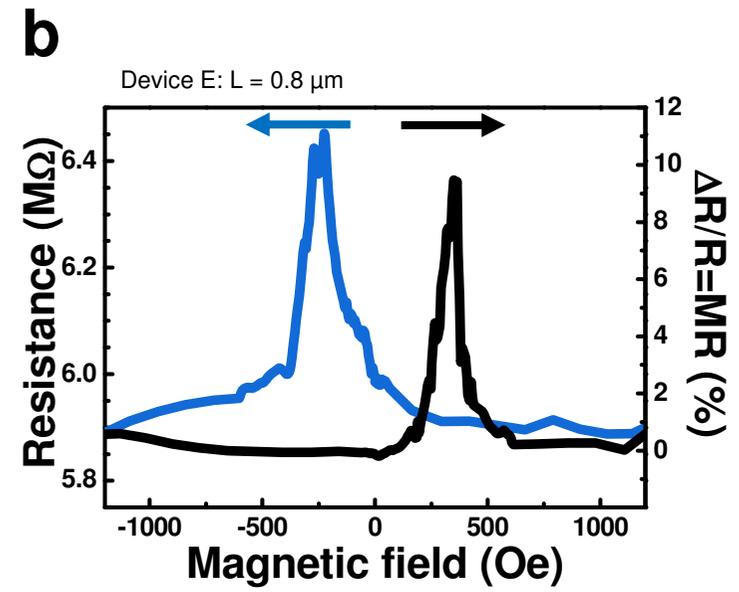

**Figure 2**

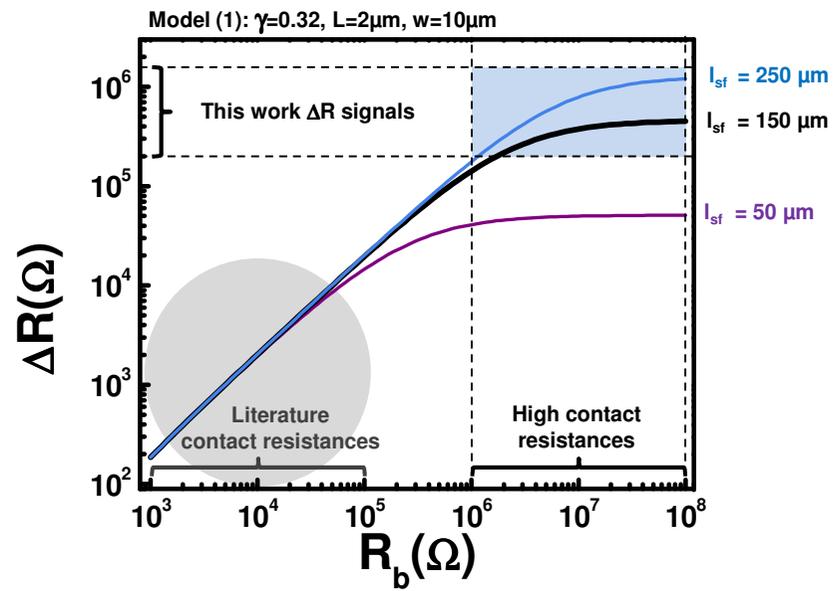

Figure 3

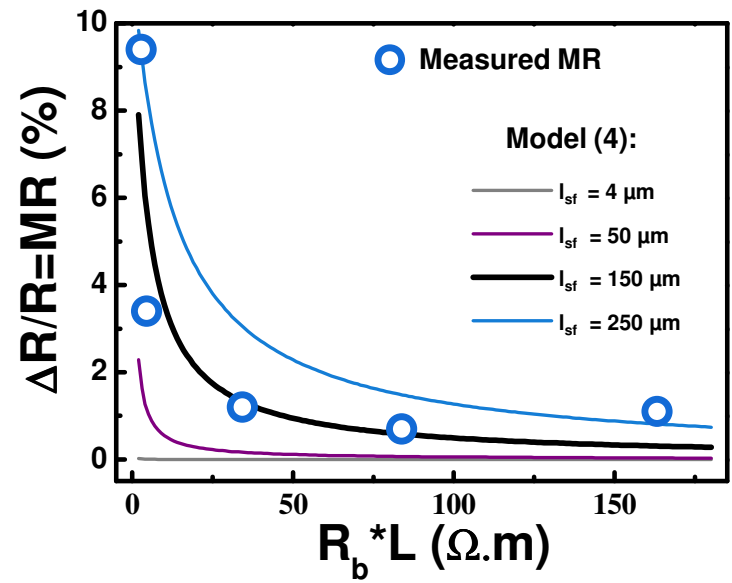

**Figure 4**